\documentclass[aps,prl,showpacs,showkeys,twocolumn,superscriptaddress]{revtex4}
\usepackage{amsmath}
\usepackage{amssymb}
\usepackage{graphicx}
\begin{document}
\title{Dielectrophoresis model for the colossal electroresistance of phase-separated manganites}
\author{Shuai Dong}
\affiliation{Nanjing National Laboratory of Microstructures, Nanjing University, Nanjing 210093, China}
\author{Han Zhu}
\affiliation{Department of Physics, Princeton University, Princeton, New Jersey 08544, USA}
\author{J.-M. Liu}
\affiliation{Nanjing National Laboratory of Microstructures, Nanjing University, Nanjing 210093, China}
\affiliation{International Center for Materials Physics, Chinese Academy of Sciences, Shenyang 110016, China}
\date{\today}

\begin{abstract}
We propose a dielectrophoresis model for phase-separated manganites. Without increase of the fraction of metallic phase, an insulator-metal transition occurs when a uniform electric field applied across the system exceeds a threshold value. Driven by the dielectrophoretic force, the metallic clusters reconfigure themselves into stripes along the direction of electric field, leading to the filamentous percolation. This process, which is time-dependent, irreversible and anisotropic, is a probable origin of the colossal electroresistance in manganites.
\end{abstract}
\maketitle

\textit{Introduction.} Phase separation (PS) can be found in various condensed matter systems. In the past half a century, the effects of electric field on phase-separated liquids have been cross-fertilized by physicists, chemists and biologists \cite{1-3,4,5,6,7}. Among them, a distinct phenomenon is dielectrophoresis \cite{8}: the mixing or demixing of neutral materials with different dielectric constants under a \textit{non-uniform} electric field \cite{4,5,6,7}. It has also found wide application in microbiology, nanotechnology, polymer processing, etc. \cite{4,9-12}. However, dielectrophoresis would seem unlikely in solids, which normally lack the necessary condition of flowability.

As for solids, PS has been demonstrated to be crucial to understanding various exotic phenomena in strongly correlated electronic materials \cite{13}, manganites being a typical example. In manganites, PS generally consists of ferromagnetic (FM) metallic phase and insulating phase which is usually antiferromagnetic and charge-ordered (CO) \cite{15}. One can undertake to understand the colossal electroresistance (CER), first observed in Pr$_{0.7}$Ca$_{0.3}$MnO$_{3}$ \cite{16}, where an applied electric field could reduce the resistance enormously. Subsequent experiments found a simultaneous uprush of magnetization, implying that the CER might be caused by the collapse of CO phase to FM phase \cite{17-18}. Although this CO-FM collapse scenario can lead to the CER, the estimated threshold of electric field is one order of magnitude larger than the experiment \cite{19}. Moreover, in a recent intriguing experiment on La$_{0.225}$Pr$_{0.4}$Ca$_{0.375}$MnO$_{3}$, no appreciable change of the magnetization was detected when the electric field induced a resistance drop of as large as four orders of magnitude \cite{20}. This anomalous CER calls for a further understanding of PS beyond the equilibrium ground state in manganites.

In this Letter, we propose a dielectrophoresis model to understand the PS dynamics associated with the CER in manganites. At the beginning, it is essential to demonstrate the possibility of dielectrophoresis in manganites being solid. Unlike liquid mixtures, the different phases in a prototype phase-separated manganite have the same chemical composition. The migration of $e_{g}$ electrons rather than cations makes the different phases in manganites 'flowable' in a wide temperature range \cite{21-22,23,24}, satisfying the first condition of dielectrophoresis: ambulatory phases. The second condition, non-uniform electric field, can be solely self-generated by the PS, as shown in Fig. 1, without external control which is necessary for dielectrophoresis in liquids \cite{9-12}. Therefore, dielectrophoresis is not impossible in manganites although they are solids.

\textit{The model.} As shown in Fig. 1, our model is based on a two-dimensional square lattice ($L\times L$, $L=64$) with a voltage drop $V$ applied in the \textit{y} direction. A fraction $p_{M}$ of the lattice sites are metallic, the others insulating. This binary mixture picture is adequate to describe the large scale PS consisting of FM and CO phases \cite{15}. A standard Metropolis algorithm is employed in our Monte Carlo simulation with the dielectrophoresis mechanism: exchange between nearest neighboring (NN) sites with the probability determined by the free energy change. Considering the free energy of electric field $F_{E}$ in the medium \cite{5,6}, the total free energy $F$ in our model can be written as,
\begin{equation}
F=F_{E}+F_{S}=-\frac{1}{2}\int\varepsilon(\textbf{\textit{r}})\textbf{\textit{E}}^{2}(\textbf{\textit{r}})d\textbf{\textit{r}}+\oint AdS,
\end{equation}
where $\varepsilon(\textbf{\textit{r}})$ denotes the dielectric constant and $\textbf{\textit{E}}(\textbf{\textit{r}})$ is the local field at point \textbf{\textit{r}}. $\varepsilon(\textbf{\textit{r}})$ equals $\varepsilon_{M}$ ($\varepsilon_{I}$) if the phase at \textbf{\textit{r}} is metallic (insulating), where $\varepsilon_{M}$ ($\varepsilon_{I}$) is the dielectric constant of the metallic (insulating) phase. In real manganites, many factors can play the role of a barrier in the transition between metastable states, e.g. domain interfaces and defects \cite{24,27}, and here we simplify the physics by characterizing them with the interface energy $F_{S}$ \cite{25-26} in Eq. (1). $F_{S}$ is proportional to the interface area $S$ between metallic and insulating regions with a coefficient $A$. Within the system, $\textbf{\textit{E}}(\textbf{\textit{r}})$ is obtained using the resistor-network (RN) model \cite{28-29,30}. In the RN model, three types of resistors are distributed according to the links between NN sites: $R_{M}$ between metallic sites, $R_{I}$ between insulating sites, and $R_{MI}$ between a metallic and an insulating site \cite{30}, set as $(R_{M}+R_{I})/2$ here. The voltage of each site $V(\textbf{\textit{r}})$ in the RN can be exactly solved by Kirchhoff equations, as shown in Fig. 1.

\begin{figure}
\includegraphics[width=220pt]{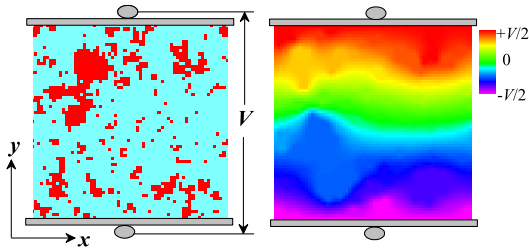}
\caption{(Color Online) Left: the phase separated lattice consisted by metallic (red) and insulating (cyan) sites. Right: the corresponding color image map of $V(\textit{\textbf{r}})$.}
\end{figure}

Our simulation begins with a zero-field quench preprocess from a high temperature $T_{0}$ to $T_{d}$. Then at $T_{d}$ fixed, the voltage $V$ changes between zero and a maximum value $V_{m}$, following an increasing-decreasing-increasing (IDI) path with a fixed rate (characterized by Monte Carlo steps (MCS)). Given the strong contrast between the electric properties of FM and CO phases \cite{15,31}, both the ratios $R_{I}/R_{M}$ and $\varepsilon_{M}/\varepsilon_{I}$ are quite large. Their values strongly depend on material, as well as doping and temperature. Experimentally, they can be determined, by comparing the metal-rich limit and insulator-rich limit \cite{15,31}. To accelerate our simulation, a special case: $f=R_{I}/R_{M}=\varepsilon_{M}/\varepsilon_{I}$ ($f\gg1$) will be used. More general cases without such restriction have also been tested. It is found that the physical picture is unaffected as long as $f\gg1$. The values of parameters used in our simulations are listed in Table. I, if not noted explicitly.

\begin{table}[b]
\caption{\textbf{Default values of parameters in the model.} Here $p_{M}$ is a little larger than the experimental value ($11-14.5\%$) in Ref. \cite{20} because the percolation threshold in 2D system is larger than that in 3D materials. The value of $f$ is within the experimental range for manganites \cite{15,31}. The other parameters are relative values.}
\begin{tabular*}{220pt}{@{\extracolsep{\fill}}lllllllr}
\hline \hline
$p_{M}$ & $\varepsilon_{I}$ & $T_{0}$ & $T_{d}$ & $A$ & MCS & $f$\\
$20\%$ & $2$ & $1.2$ & $1$ & $0.85$ & $1.5\times10^{3}$ & $3\times10^{3}$\\
\hline \hline
\end{tabular*}
\end{table}

\textit{Results.} The resistance $R_{y}$ along the direction of the electric field is shown in Fig. 2 as a function of the external electric field $E$ ($E=V/(L-1)$). Initially, the resistance is only weakly dependent on $E$. Then, the CER transition takes place once $E$ exceeds a threshold $E_{c}$. In the subsequent stages of increasing/decreasing $E$, $R_{y}$ remains low and nonlinear even for $E$ well below $E_{c}$. We find that the value of $E_{c}$ increases with the IDI rate (not shown here), in agreement with the observations in La$_{0.225}$Pr$_{0.4}$Ca$_{0.375}$MnO$_{3}$ \cite{20}, implying that the process is a relaxative evolution between metastable states. We use the ratio $\delta_{ER}=R_{y}^{MAX}/R_{y}^{MIN}$ to characterize the CER, where $R_{y}^{MAX}$ ($R_{y}^{MIN}$) is the maximum (minimum) of $R_{y}$ before (after) the CER transition. $\delta_{ER}$ appears to be proportional to $f$, as shown in the inset of Fig. 2.

\begin{figure}
\includegraphics[width=220pt]{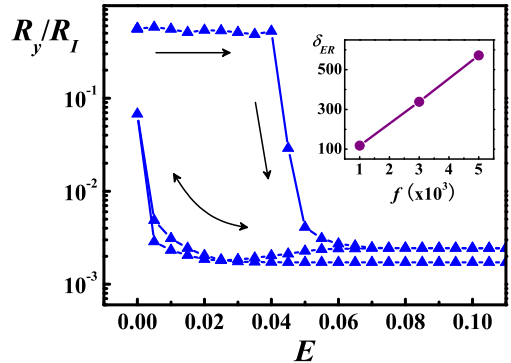}
\caption{(Color online) Resistances as a function of the applied field. Inset: $\delta_{ER}$ as a function of $f$.}
\end{figure}

To understand the above transport behavior, four snapshots of PS pattern taken at different stages during the IDI process are shown in Fig. 3. Before $E$ exceeds $E_{c}$, as shown in Fig. 3(a), the metallic sites accumulate into clusters in favor of lower interface energy. However, once $E>E_{c}$, the energy gained from dielectrophoresis overcomes the interface energy barrier, and the PS structure is rapidly reconfigured into a stripe-like pattern, as shown in Fig. 3(b). As a result of the percolation, the system has higher overall conductance and higher effective dielectric constant. Upon further increasing the field, the percolation path is strengthened, as shown in Fig. 3(c), giving rise to a slight decrease of resistance. When the field is decreasing, the percolation path is hardly destroyed except when $E$ is close to zero. When $E$ is too small to compete against the surface tension and thermal energy, the percolation path will be corroded gradually and, given enough time, eventually break down, as shown in Fig. 3(d). Nevertheless, with the parameters as listed in Table I, the main shape of percolation path can remain effective for a finite period of time, which accounts for the nonlinear transport behavior observed in Fig. 2 after the voltage threshold is first exceeded. As a result, the dielectrophoresis in our simulation is irreversible.

\begin{figure}
 \includegraphics[width=220pt]{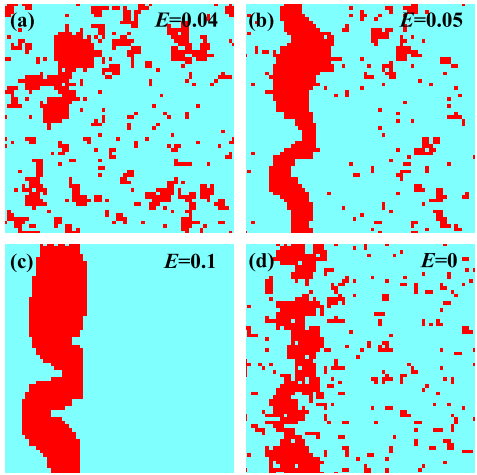}
\caption{(Color Online) Four snapshots taken from an evolution cycle of dielectrophoresis.}
\end{figure} 

Besides the changing rate of the electric field, the interface energy coefficient $A$ and temperature also play important roles in modulating the dielectrophoresis. In our simulation, the dependence of ($E_{c}/A$) upon ($T_{d}/A$) is quite nontrivial, as shown in Fig. 4. ($E_{c}/A$) reaches the minimum at ($T_{d}/A$) $\sim1.1$. It agrees with the experimental evidences that there is a $T$-window for the dynamic PS ($T\sim25-85$ K) \cite{23} or a novel consolute critical point $T\sim30$ K \cite{20} in the phase-separated La$_{0.225}$Pr$_{0.4}$Ca$_{0.375}$MnO$_{3}$. The origin could be that, at large ($T_{d}/A$), the thermal energy can destroy a stable percolation of the metallic phases; on the other hand, at small ($T_{d}/A$), the relatively large surface tension can effectively prevent the metallic clusters from being reshaped into stripes. Therefore, ($E_{c}/A$) is larger in both cases of large and small ($T_{d}/A$). Although the parameter $A$ can not be directly measured in experiments, the consolute critical point does provide a way to compare our result of $E_{c}$ with experimental results. Using the parameters¡¯ values taken from experiments \cite{15,20,31}, $E_{c}$ in our model is estimated to be $\sim10^3$ V/m, within the same order of magnitude of the experimental data \cite{20}. As we compare the physics of the present filamentous percolation picture and the CO-FM collapse picture, one notable thing is that $E_{c}$ ($\sim10^6-10^7$ V/m \cite{16,19}) required for the CO-FM collapse is indeed much larger than $E_{c}$ in our model. The reason is that, the CER here is induced by merely repatterning of the CO and FM phases, not by converting them from one to the other, which costs more energy to compensate for the energy difference between CO and FM phases.

\begin{figure}
 \includegraphics[width=220pt]{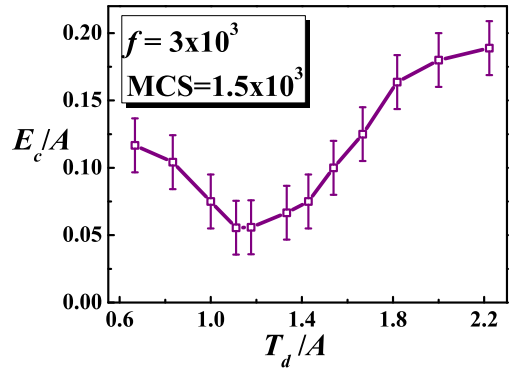}
\caption{(Color online) The field threshold as a function of the temperature. Here the surface tension coefficient $A$ is chosen as the unit while the other parameters have the values in Table I.}
\end{figure} 

Since the metallic filaments along the direction of electric field break the space symmetry of PS, it is natural to expect transport anisotropy. In our simulation, resistances both along ($R_{y}$) and across ($R_{x}$) the electric field are affected, as shown in Fig. 5. Before the voltage first reaches the threshold, the transport is isotropic, and resistances along both $x$ and $y$ are highly independent of the voltage drop. At the voltage threshold, the CER transition is accompanied by the sharp emergence of transport anisotropy. While $R_{y}$ along the direction of the electric field is drastically reduced, a simultaneous increase of $R_{x}$ across the electric field occurs irreversibly near the threshold voltage. Furthermore, the system has the intriguing feature that it is metallic ($dR_{y}/dT>0$) along the electric field direction, but insulating ($dR_{x}/dT<0$) in the perpendicular direction.

\begin{figure}[b]
 \includegraphics[width=220pt]{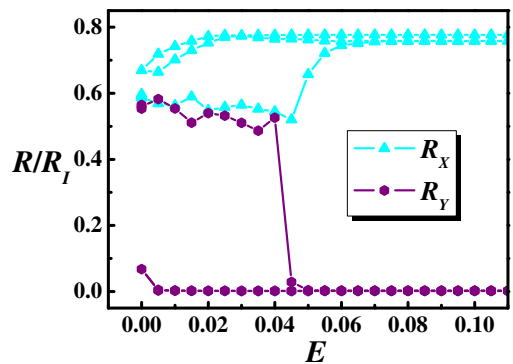}
\caption{(Color online) Comparison of the perpendicular (cyan) and parallel (purple) resistances as a function of the applied field.}
\end{figure} 

\textit{Discussion.} Using the Bruggeman-Landauer (BL) equation based on self-consistent effective medium approximation (SCEMA) \cite{33}, the effective conductance of a well-distributed and isotropic binary mixtures system can be estimated as,
\begin{equation}
C=\frac{1}{2(d-1)}(-P+\sqrt{P^{2}+4(d-1)C_{I}C_{M}}),
\end{equation}
where $d$ is the dimensionality and $C_{I}=1/R_{I}$ ($C_{M}=1/R_{M}$) is the conductance of insulting (metallic) phases; $P=C_{M}(1-dp_{M})+C_{I}(dp_{M}+1-d)$. Since $C_{I}/C_{M}\sim0$, Eq. (2) can be simplified to $C\approx C_{I}/(1-dp_{M})$ for all cases ($f\gg1$) in our model. Therefore, the effective resistance is about $(1-dp_{M})R_{I}=0.6R_{I}$ for all values of $f$ considered here and the conductance is isotropic. In our simulation, the conductance is isotropic and its value is indeed independent of $f$ before the CER transition. The value of $R_{y}\approx0.55R_{I}$ is close to, but slightly smaller than, the BL estimation. The reason is that the accumulation of metallic phases into clusters will enhance the conductance. As the percolation path is formed, the $R_{y}^{MIN}$ can be estimated as $R_{I}/(p_{M}f+1-p_{M})\approx R_{I}/(p_{M}f)$ since $f\gg1$. It accounts for the linear dependence of $\delta_{ER}$ on $f$, as shown in the inset of Fig. 2. (The relationship $\delta_{ER}\propto R_{I}/R_{M}$ is independent of the choice of $\varepsilon_{M}/\varepsilon_{I}$ in our model.) On the other hand, $R_{x}^{MAX}$ after the CER transition is approximately $R_{I}(1-p_{M})+R_{M}p_{M}\approx R_{I}(1-p_{M})$. Compared with its original value, there is a relative small change (increase) of $R_{x}$: $R_{x}^{MAX}/R_{x}^{MIN}=(1-p_{M})/(1-2p_{M})$ by the two-dimensional BL equation.

Compared with the dielectrophoresis in liquids, there are two remarkable features in phase-separated manganites, adding to its appeal to the research community. Firstly, the prototype PS in manganites would generate inhomogeneous electric field by itself, while for liquids great effort was made to artificially produce non-uniform electric field \cite{6,9-12}. Yet, as in the cases of the liquids, the PS in manganites can also be controlled by applying artificially non-uniform electric field (e.g., patterning of gates), offering great potential for making devices. Secondly, the diversity of electric properties between different phases in manganites, e.g. resistance and dielectric constant, is far greater than those in consolute liquids. In real manganites, the colossal electroresistance/magnetoresistance implies a colossal difference between the resistivities of metallic and insulating phases. Therefore, the main difficulties of dielectrophoresis in liquid, i.e. a large voltage required and weak effects, could be overcome in manganites.

In conclusion, we introduced the dielectrophoresis dynamics, a phenomenon usually existing in mixed phase liquids, to investigate the phase separation in solid manganites. We have shown that the colossal electroresistance could be achieved by the filamentous percolation, through a dielectrophoresis dynamic process without increasing the metallic composition. Our results give a probable explanation of the colossal electroresistance in manganites, including its time-dependence, irreversibility and the consolute critical point. Besides, it was also predicted that the dielectrophoresis process renders the system anisotropic. All these properties can not be explained by the physics of single phases but arise from phase competition. The present work also highlights the analogy \cite{13} between strongly correlated electronic systems and complex systems in the classical world, e.g., complex fluids.

We thank Dr. Carlos Acha for helpful discussions. This work was supported by the Natural Science Foundation of China (50332020, 10021001) and National Key Projects for Basic Research of China (2006CB0L1002). S. Dong was supported by the Scientific Research Foundation of Graduate School of Nanjing University (2006CL1).

\end{document}